\documentclass[notitlepage,aps,prd,amsmath,floats,floatfix,twocolumn,superscriptaddress,nofootinbib,showpacs]{revtex4}

\usepackage{amsfonts,amsmath,wasysym,epsfig,graphicx,verbatim,color,subfigure,graphicx}
\graphicspath{{Images/}}

\usepackage[range-phrase=--,range-units=single,retain-unity-mantissa=false]{siunitx}
\usepackage{amsmath}
\usepackage{amssymb}
\usepackage{amsfonts}
\usepackage{bm}
\usepackage{color}
\usepackage[normalem]{ulem}
\usepackage{xcolor}
\usepackage{natbib}
\usepackage{color,soul}
\usepackage{hyperref}
\usepackage{marginnote}
\usepackage{braket}
\usepackage[none]{hyphenat}
\usepackage{bm}
\usepackage{comment}
\usepackage{multirow}
\usepackage{xspace}
\usepackage[toc,page]{appendix}
\usepackage{float} 

\newcommand{\silica}{SiO$_{2}$\xspace}
\newcommand{\tantala}{Ta$_{2}$O$_{5}$\xspace}

\ifdefined\COMMENTS
    
    \newcommand{\red}[1]{{\color{red}#1}}
    \newcommand{\blue}[1]{{\color{blue}#1}}

\else
    
    \newcommand{\red}[1]{{#1}}
    \newcommand{\blue}[1]{{#1}}

\fi

\newcommand{\SU}{\affiliation{Department of Physics, Syracuse University, Syracuse, New York 13244, USA}}

\newcommand{\cardiff}{\affiliation{School of Physics and Astronomy, Cardiff University, Cardiff CF24 3AA, United Kingdom}}

\date{\today}

\begin{document}

\title{Birefringence of AlGaAs/GaAs Coatings under Above-Band-Gap Illumination, \newline GR Noise and Photo-Optic Transfer Function}

\author{Bin Wu}\email{bwu127@syr.edu}\SU
\author{Shreyan Goswami}\SU
\author{Satoshi Tanioka}
\altaffiliation[Present address: ]{Quantum and Spacetime Research Institute, Kyushu University, Nishi-Ku, Fukuoka, 819-0395, Japan.}
\cardiff
\author{Stefan Ballmer}\SU

\begin{abstract}
AlGaAs/GaAs coatings are being considered as coating candidates for gravitational-wave detectors. In this paper we investigate the birefringence properties of this crystalline semiconductor material by modulating the optical illumination on the mirror coating and monitoring the induced birefringence. While the measured low-frequency birefringence values align with previous studies, we observed a frequency-dependent behavior in the illumination-to-birefringence coupling, characterized by a pole increasing with illumination intensity and a gain at zero frequency (DC gain) decreasing with illumination intensity. We developed a generic theoretical model based on a master equation to characterize the measurement results by considering photon-induced electric fields and electro-optic effects. This model can fit the frequency and intensity dependencies of the induced birefringence. Additionally, this model predicts a generation-recombination noise (GR noise) will be observable in the coating birefringence. While the presented measurement cannot predict the exact level of GR noise, for the frequency band and spot sizes relevant for gravitational-wave detectors we expect GR noise to be white below the pole frequency, scale with power the same way laser shot noise does, and for fixed power be independent of spot size.

\end{abstract}

\maketitle
\section{Introduction}\label{sec:intro}
Currently, the sensitivity of ground-based gravitational-wave detectors is limited by thermal noise from optical test mass coatings at around 100 Hz~\cite{O4_sensitivity,ctn_2024,evans_2008}. The fluctuation-dissipation theorem directly connects this thermal noise to the mechanical dissipation~\cite{Kubo_1966, Levin_1998} and thermal dissipation~\cite{PhysRevD.91.023010,evans_2008,Chalermsongsak:2015cya} in the optical coatings. Today, these coatings are dielectric stacks of sputtered amorphous materials, alternating between the Ti-doped \tantala and \silica~\cite{harry_2007,Amato_2019}. 

A very promising alternative are AlGaAs/GaAs crystalline coatings, which exhibit extremely low intrinsic mechanical dissipation \cite{cole08, Cole2013,Cole:16}. Indeed, the lowest thermal noise reference cavities to date use this coating~\cite{lee2025frequency}.
In these crystalline semiconductor coatings electric field fluctuations cause a birefringence, i.e., a difference in the reflected field phase between s- and p-polarization, which mimics as mirror motion for a single polarization~\cite{EO_effect,Tanioka_2023}. Moreover, the  
semiconductor band structure implies that under high optical intensities charge carriers can be excited. Even for photon energies below the material band gap this can still happen via multi-photon processes. Also band-bending at the coating interface structure can lead to trapping of charge carriers \blue{in} the coating~\cite{doi:10.1021/cr3000626}.
This can result in changes in the internal electric field across the coating, and therefore changes in the phase of light reflected off the coating. Since the effect has opposite sign for p- and s-polarization, this coupling can be measured by comparing the resonant frequency of the two polarizations in a cavity\red{~\cite{Tanioka_2023,Yu2023}}.

According to \blue{results from groups at the Physikalisch-Technische Bundesanstalt and the Joint Institute for Laboratory Astrophysics} (PTB and JILA), the static birefringence in AlGaAs coating is $n_{\rm biref}\approx7\times 10^{-4}$~\cite{Yu2023}, where $n_{\rm biref} $ refers to the static refractive index difference between the two polarizations. They also observed that $n_{\rm biref}$ depends on the intracavity light intensity. Similar results are obtained when the coatings are illuminated with above-band-gap photons~\cite{Ma_2024}.

In this paper we study the birefringence change, \blue{i.e., the change in the cavity frequency splitting between p- and s-polarization,} introduced by light-induced charge carriers in the semiconductor coating. We built an optical cavity with one $\rm Al_{0.92}Ga_{0.08}As/GaAs$ mirror and measured the transfer functions from the above-band-gap illumination to the birefringent line splitting between two orthogonal polarizations under different DC powers. In addition, to study the wavelength dependence of external illumination, we used two different light sources: 700 nm (higher than the band gap of GaAs but lower than that of AlGaAs) and 430 nm (higher than the band gaps of both GaAs and AlGaAs), and varied the intra-cavity carrier (1064 nm) intensity.
\blue{We observe that the measured transfer function exhibits a low-pass behavior with a pole at a few hundred Hz. The DC gain of that transfer function depends on the illumination intensity, consistent with previous results~\cite{Ma_2024}. Notably though the pole frequency is also dependent on the illumination intensity, which rules out a thermal diffusion or photo-elastic origin as the reason for the low-pass filter~\cite{PhysRevD.91.023010,DeRosa2002}. In appendix \ref{sec:MASTERTheory}, we provide a master equation model that can describe the observed intensity-dependent behavior by assuming that photo-excited charges get temporarily trapped somewhere in the coating. Since we want a generic framework to describe the experimental data, we intentionally avoid any assumptions about the microscopic details of this trapping mechanism.}
\blue{Still, this} model predicts the existence of generation-recombination (GR) noise. We derive the power spectrum density of the birefringence GR noise, which also gives us the noise scaling with optical power and spot size.


\section{Experiment design}\label{sec:sec2}

Birefringence results in a frequency splitting ($f_{\rm biref}$) between the linearly polarized fast and slow eigenmodes of an optical cavity. The frequency splitting is proportional to the birefringence angle ($\phi_{\mathrm{biref}}$) of the coating,  and to the cavity free spectral range, ${\rm FSR} = \frac{c}{2L}$, where $c$ denotes the speed of light and $L$ is the cavity length:

\begin{equation}
    f_{\mathrm{biref}}= {\mathrm {FSR}} \times \frac{\phi_{\mathrm{biref}}}{2\pi}.
\end{equation}

\blue{$\phi_{\mathrm{biref}}$ is defined as the phase difference acquired
between the p- and s-polarizations upon reflection from the coating.}
Thus the shorter the cavity, the larger the frequency splitting. 
But a short cavity also results in a small beam spot and therefore a comparatively high 1064~nm beam intensity. We wanted to keep the intensity below about $\rm 5 MW/m^2$ to reduce the impact of below-band-gap light and chose the input laser power and cavity parameters listed in Table~\ref{tab:Cav_parameeters} accordingly. The output coupler of our cavity is an $\rm Al_{0.92}Ga_{0.08}As/GaAs$ mirror with 81 aperiodic layers and is optimized to minimize thermo-optic noise for 1064 nm~\cite{Chalermsongsak:2015cya}. While varying the aluminum-to-gallium ratio can yield different electrical and optical properties, the coating we used has been demonstrated to exhibit low mechanical loss and high optical quality, meeting the standard requirements for gravitational-wave detectors~\cite{Cole:16}. We measured a static frequency splitting of about \SI{1}{MHz}, corresponding to a birefringence angle $\phi_{\mathrm{biref}} = 1.7\times 10^{-3}$, within the range of literature values \cite{ctn_2024}.

\begin{table}[ht]
\centering
\caption{\bf Cavity and coating parameters.}
\begin{tabular}{ll}
\hline
Parameter & Value \\
\hline
Cavity length & $4\,\unit{cm}$ \\
Free spectral range & $3.75\,\unit{GHz}$ \\
Finesse (measured) & $\sim500$ \\
FWHM linewidth & $7.8\, \unit{MHz}$ \\
Beam spot $w$ on coating & 0.3 \unit{mm}\\
{1064 nm intensity range}\;\;& $2.0 - 3.7\,\unit{MW/m^2}$\\
Input coupler reflectivity & $98.8\%$ at $1064\,\unit{nm}$ \\
Output coupler reflectivity\\
($\rm Al_{0.92}Ga_{0.08}As/GaAs$ coated)& $99.99\%$ at $1064\,\unit{nm}$ \\

\hline
\end{tabular}
  \label{tab:Cav_parameeters}
\end{table}

\begin{figure}
\begin{center}
\includegraphics[height=0.18\textheight]{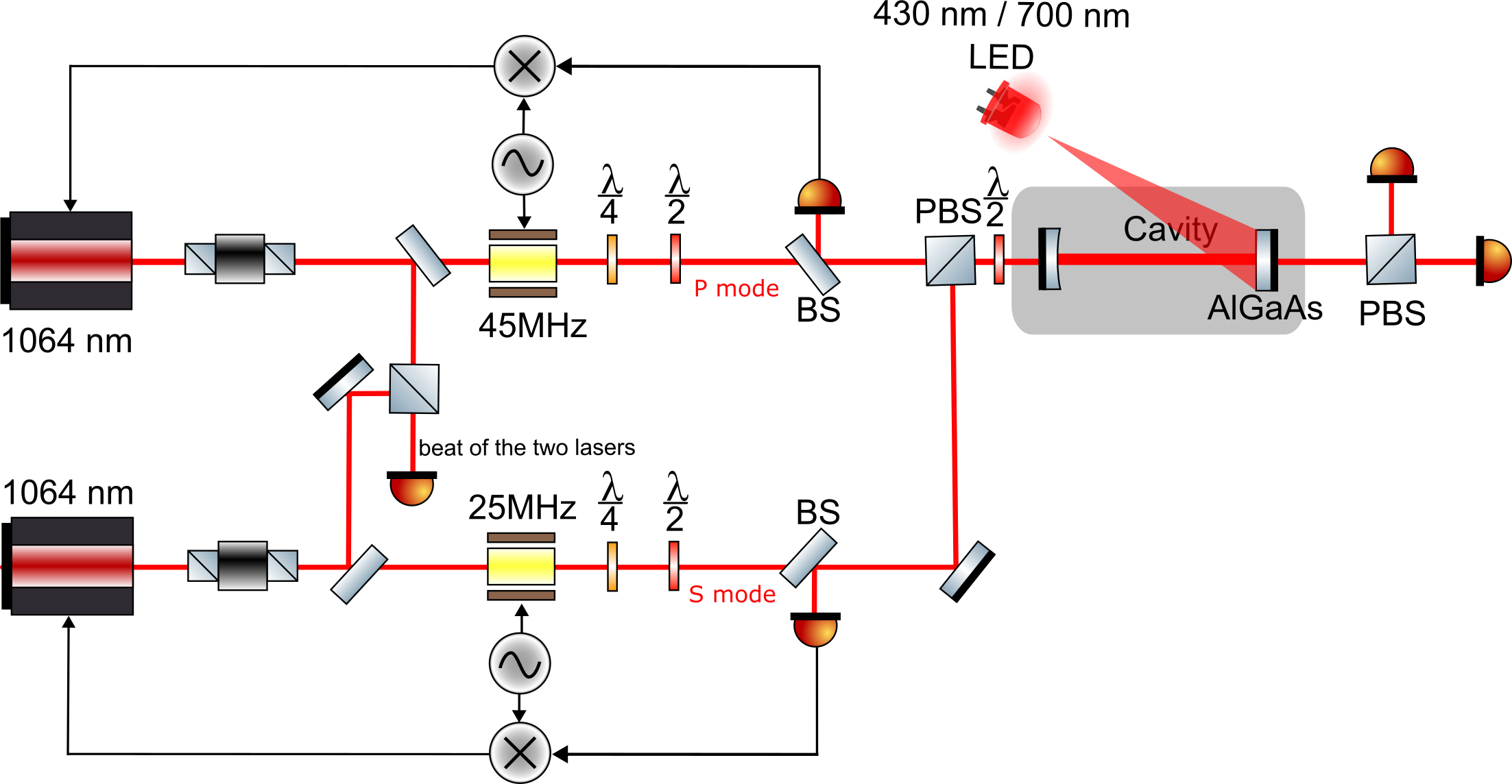}
\caption{Experiment Scheme. The optical cavity is 4-cm long, with its output coupler coated with $\rm Al_{0.92}Ga_{0.08}As/GaAs$. To monitor the frequency splitting between orthogonal polarizations induced by the illumination, we measured the beat note between two independent lasers, one locked to the cavity in p-polarization and the other in s-polarization.}
\label{fig:setup}
\end{center}
\end{figure}

The setup of the experiment is shown in FIG.~\ref{fig:setup}. Two lasers with different polarizations are locked to a 4-cm short cavity simultaneously using PDH locking \cite{pdh} with sidebands at \SI{25}{MHz} and \SI{45}{MHz} respectively.  A half-wave plate is placed right before the cavity to rotate the p- and s-modes to align them with the fast and slow axes of the crystalline coating. Additionally, the beat note between the two lasers is measured by a photodiode, and either directly recorded at \SI{10}{MHz} sampling rate, or, to simplify transfer function measurements, fed into a phase-locked loop. The control signal of that phase-locked loop is a direct measurement of the introduced frequency shift. 

\blue{The AlGaAs mirror is illuminated with an LED at 700~nm, as well as 430~nm wavelength for comparison. The 700~nm LED's intensity is driven from 1 Hz to 10 kHz with a DC offset ranging from 0.1 $\rm mW$ to 0.7 $\rm mW$ (0.6 $\rm W/m^2$ to 5 $\rm W/m^2$ at the sample) and a modulation amplitude of 0.06 $\rm mW$ (0.4 $\rm W/m^2$ at the sample), corresponding to a modulation index range of $0.08$ to $0.67$. The minimum and maximum DC offsets of the driver are limited by the linear operation range of the LEDs and the maximum output voltage of the voltage amplifier. 
We chose a high modulation index as this provides a better signal-to-noise ratio for the LED-intensity to birefringence-frequency-split transfer function. We verified that decreasing the modulation index had no systematic impact on the transfer function other than increasing the measurement noise.}
 
The amplitude of the introduced beat note frequency shift $\Delta f_{\mathrm{induced}}$ is given by
\begin{equation}
    \Delta f_{\mathrm{induced}} =  {\rm FSR} \times \frac{\Delta \phi_{\mathrm{biref}}}{2\pi},
\end{equation}
where $\Delta \phi_{\mathrm{biref}}$ denotes the illumination-induced birefringent angle.

\section{Measurement results}\label{sec:res}
A phase-locked loop (PLL) was implemented to track the beat note between the two lasers. The coupling coefficients from illumination to coating birefringence were determined from the transfer function between the LED drive and the PLL control signal, followed by a series of calibrations. This section presents the main measurement results for different illumination conditions. Details of the measurement scheme and calibration are provided in Appendix~\ref{sec:scheme}.

\subsection{Frequency dependence}
\label{subsec:freqdep}

\begin{figure}
\begin{center}
\includegraphics[width=0.4\textwidth]{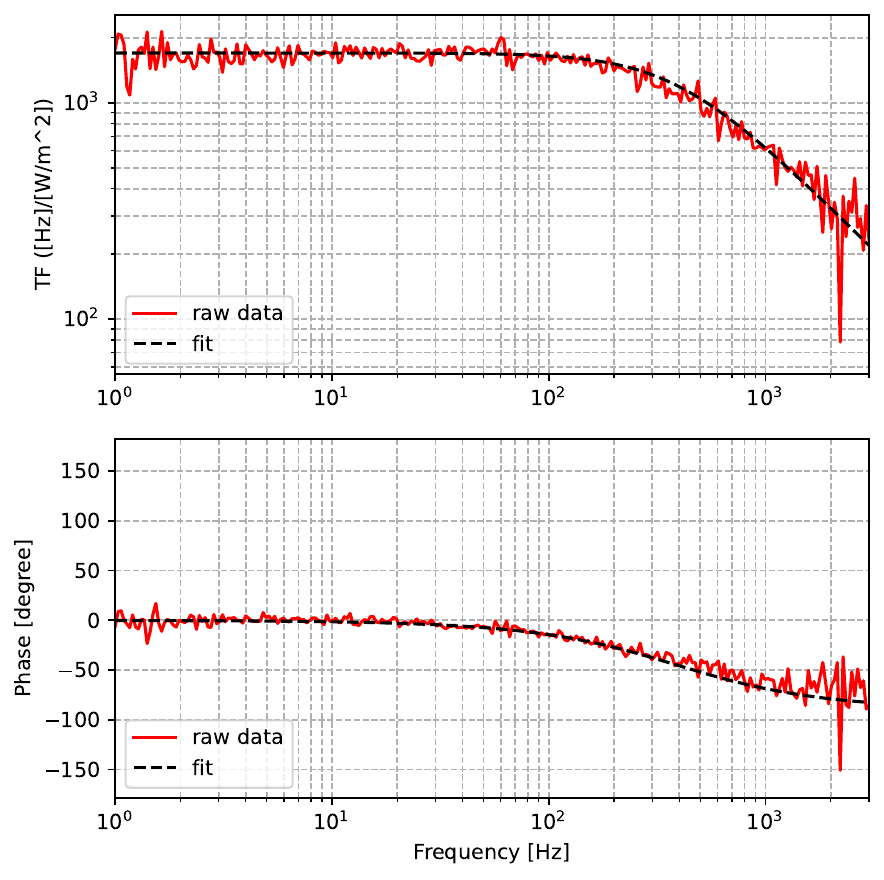}
\caption{Typical transfer function, \SI{1}{Hz} to \SI{3}{kHz}, from LED illumination to beat note frequency. The solid line shows the raw data and the dashed line is a fitting to a first-order low-pass filter. 1064 nm intensity: $2.0\,\rm MW/m^2$, LED wavelength: 700 nm.
\blue{Note that the drive electronics and LED light angle-of-incidence for this measurement was different from those for FIG. 3, making a comparison difficult.}
}
\label{fig:TF_result1}
\end{center}
\end{figure}

FIG.~\ref{fig:TF_result1} shows a typical transfer function \blue{describing the coupling 
between the intensity-modulated optical pump and the coating birefringence. It} indicates that the coupling coefficient from the illumination to the induced birefringence is frequency dependent. Here, the LED wavelength is 700 nm and the intensity is 1.5 $\rm W/m^2$ with a modulation of $1\; \rm W/m^2$, corresponding to a modulation index of $m=0.67$.
\blue{Note that we report the incident light intensities at the mirror; the exact absorbed power at the coating surface also depends on the incident angle and wavelength.} The transfer function exhibits two main features: \blue{a frequency-independent gain below the roll-off,} and a single-pole roll-off at around $300~{\rm{Hz}}$. The pole frequency suggests a $1/e$ lifetime of a few hundred microseconds for the relevant excited carriers in the coating. Both the phase and amplitude of the measurement data can be well fitted to a first-order low-pass filter. We lose coherence above around $\rm 2\, kHz$, which is largely due to insufficient gain of the laser locking loops.

The DC value of the transfer function FIG.~\ref{fig:TF_result1} is around $\SI{1.7e3}{Hz/(W/m^2)}$. We can convert this into an induced phase shift using the FSR of our cavity: 
\begin{equation}
    \Delta \phi = 2\pi \frac{\Delta f}{\rm FSR} = 2 \pi \frac{\SI{1.7e3}{Hz/(W/m^2)}}{\SI{3.75}{GHz}}  I,
\end{equation}
\begin{equation}
    \frac{\Delta \phi}{I} = \SI{2.8e-6}{rad/(W/m^2)} .
\end{equation}

\blue{This transfer function's DC value corresponds to the slope of an adiabatic  birefringence vs intensity curve, which was measured by the PTB and JILA groups in \cite{Ma_2024}, Fig. 4. The coupling coefficients derived from their results are in the range of $10^{-6}\sim10^{-5}\; \rm {rad/(W/m^2)}$, roughly comparable with our result, and decreasing with intensity. Note that their cavity carrier wavelength is 1550 nm, and the LED illumination wavelengths are different from ours.}

\subsection{Change in illumination intensity}
\begin{figure}
\begin{center}
\includegraphics[width=0.3\textheight]
{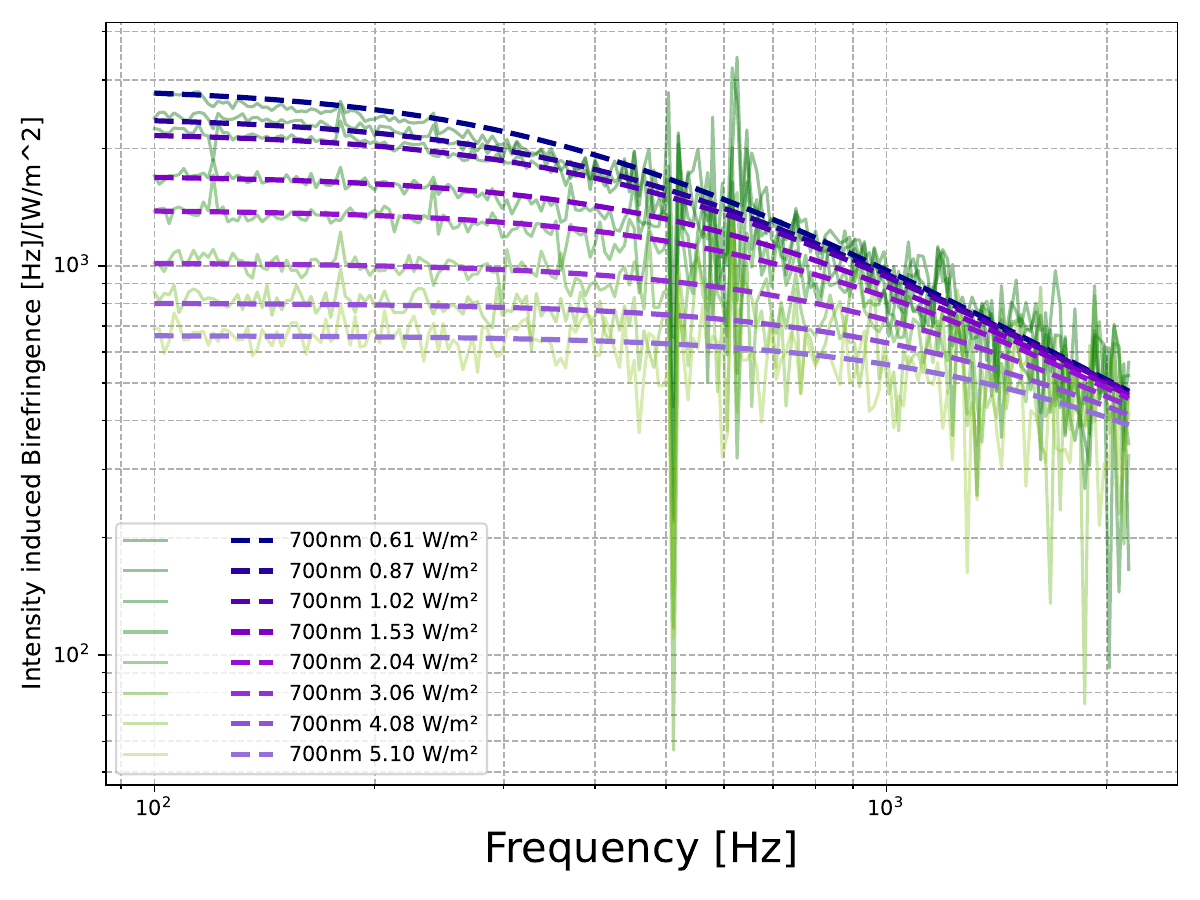}
\caption{Fitting results at different illumination intensity levels: The wavelength of the illumination is 700 nm. The solid green lines are measured raw data after calibration. The noise peaks at $\sim$ 500 Hz and $\sim$ 600 Hz are due to the mechanical resonance of the cavity mirror posts. The dashed lines are the fitting results obtained by the four-parameter global fit described in section \ref{sec:GlobalFit} and appendix \ref{sec:MASTERTheory}. This plot shows data for a \SI{1064}{nm} carrier light intensity of \SI{2.0}{MW/m^2}. See FIG.~\ref{fig:globalfit} (a) and (b) for other \SI{1064}{nm} intensities. \blue{
The drive electronics and light incident angle on the mirror is different from that used in FIG.~\ref{fig:TF_result1}.} }
\label{fig:change DC2}
\end{center}
\end{figure}

To study the effect of intensity level on the coupling coefficient, we then changed the DC power of the illumination. The results of multiple measurements are shown in FIG.~\ref{fig:change DC2}. The solid lines are measured raw data and the dashed lines are the fitting results obtained by the four-parameter global fit described in section \ref{sec:GlobalFit} and appendix \ref{sec:MASTERTheory} below. We find that as the intensity increases, the coupling coefficient decreases and the pole frequency increases. More specifically, in FIG.~\ref{fig:change DC2} as the DC intensity $I_1$ of the 700 nm LED varies from $\rm 0.6\, W/m^2$ to $\rm 5 \,W/m^2$, the DC coupling levels decrease from $2760$ to $660$ $\rm [Hz]/(W/m^2)$. Above the rising pole frequency the transfer function is independent of illumination intensity, at least within the measurement noise.

The red dots in FIG.~\ref{fig:dcgain_fit} show the DC values in FIG.~\ref{fig:change DC2}. The yellow and green dots correspond to different levels of $1064~{\rm {nm}}$ intra-cavity intensity - see FIG.~\ref{fig:globalfit} and section~\ref{ssCI}. \blue{The decrease of the DC values with intensity seen in FIG.~\ref{fig:dcgain_fit} is consistent with the decrease of the slope in Fig. 4 of \cite{Ma_2024}.}

\subsection{Change in cavity carrier intensity}\label{change1064}
\label{ssCI}
\begin{figure}
\begin{center}
\includegraphics[width=0.45\textwidth]{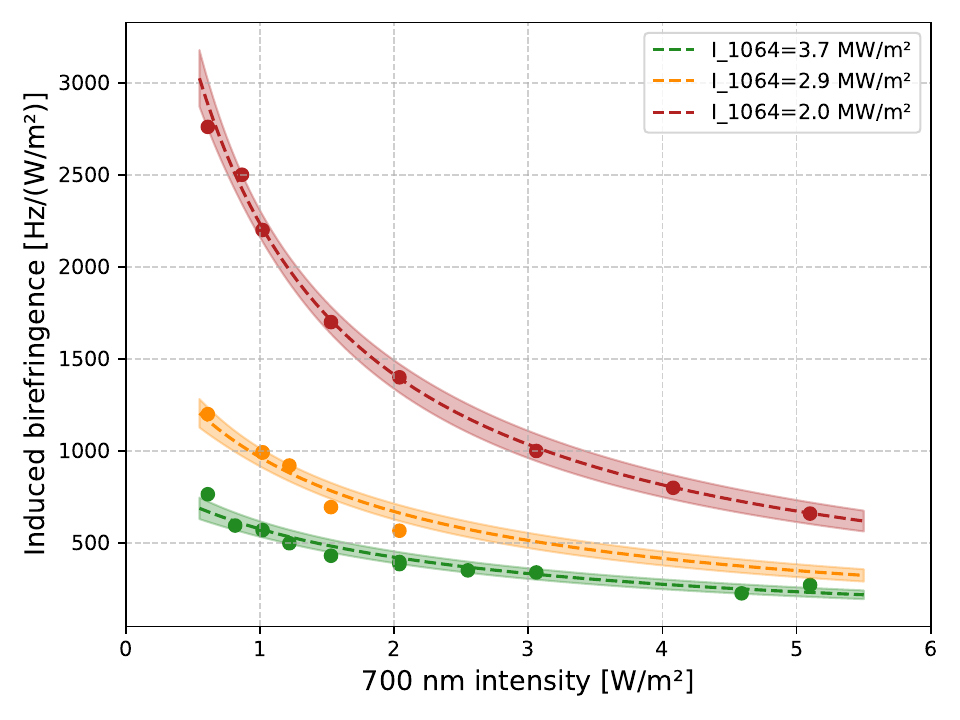}

\caption{DC gain values for the induced birefringence under various of 700~nm and 1064~nm intensities. The x-axis is the intensity of 700~nm illumination and different colors represent varying 1064~nm intensities.
The data points come from the measurements shown in FIG.~\ref{fig:change DC2} and FIG.~\ref{fig:globalfit}. The dashed lines and shaded uncertainties are calculated based on the model described in Appendix~\ref{sec:MASTERTheory}.
}\label{fig:dcgain_fit}
\end{center}
\end{figure}

The wavelength of the carrier in our experiment is 1064~nm, which is below the band gap of AlGaAs and GaAs. However, below-band-gap illumination can also excite charge carriers and thus induce birefringence through two-photon absorption on a time scale much shorter than our observations. Indeed FIG.~\ref{fig:dcgain_fit} shows that higher carrier intensities result in lower DC gain values. The dashed lines are predicted values from the global fit model described in section \ref{sec:GlobalFit} and appendix \ref{sec:MASTERTheory}, and the shaded area shows the uncertainty range.

\newpage
\subsection{Comparison between 700 nm and 430 nm illumination}
\blue{

We originally chose $700\,$nm illumination because its photon energy
(\SI{1.77}{eV}) lies above the band gap of GaAs but below that of AlGaAs.
To extend the measurement to photon energies above both band gaps, we swapped
the LED from $700\,$nm to $430\,$nm and repeated the measurement. FIG.~\ref{fig:change
wavelength} shows the results. The overall response is similar to the
$700\,$nm case---flat at low frequencies with a roll-off at higher
frequencies---but two differences stand out. First, at comparable intensities
the DC gain under $430\,$nm illumination is several times lower, which we in part attribute to the stronger absorption of GaAs at the shorter wavelength (\SI{2e4}{cm^{-1}} at $700\,$nm versus \SI{7e4}{cm^{-1}} at $430\,$nm~\cite{ioffe}): significantly less light reaches the coating interface layers.
Second, the $430\,$nm data are better described by a superposition
of pole frequencies than by a single low-pass filter, indicating a broadened
distribution of effective carrier lifetimes. Such a spread is expected if, for
example, the carrier lifetime depends significantly on the $1064\,$nm
intensity, which in turn varies across the Gaussian readout beam spot.}

Because our \SI{700}{nm} data was much cleaner, both in terms of single pole frequency and overall signal-to-noise, we did not include the \SI{430}{nm} data in the global fit presented in Table~\ref{tab:fit_parameters} and the noise models presented below.

\begin{figure}
\begin{center}
\includegraphics[height=0.25
\textheight]{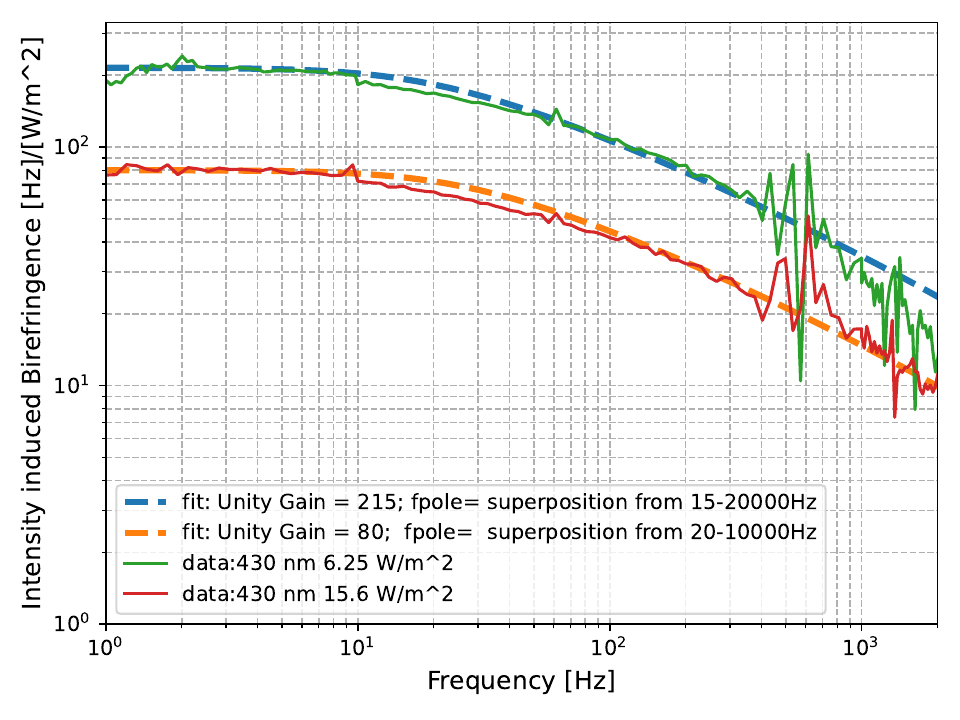}
\caption{ Transfer functions from 430 nm LED intensity to beat note noise. Solid lines represent measurement data under different illumination intensities. Dashed lines indicate fitted curves for the raw data. Each of the fitted curves here is not simply a pure first-order low-pass filter, but consists of a series of superimposed first-order low-pass filters.}
\label{fig:change wavelength}
\end{center}
\end{figure}

\section{Master Equation and Noise}
\subsection{Global Fit to Master Equation}
\label{sec:GlobalFit}

The behavior of the measurement results presented in sections~\ref{subsec:freqdep} through~\ref{ssCI} can be captured by a master equation model for the number of charge carriers $N$ across the effective area responsible for an induced charge area density $\sigma=e~N/A$ trapped \blue{somewhere in} the coating layers, which in turn causes birefringence through the electro-optic effect ~\cite{EO_effect}. Here $e$ is the elementary charge, and $A$ is the effective area. 
Details are described in appendix~\ref{sec:MASTERTheory}. The master equation up to linear order in $(N-\bar{N})$ can be written as (see equation \ref{eq:EffMaster}):
\begin{align}
    \dot{N} = -\Gamma(I_i) \left( N - \bar{N}(I_i) \right) ,
    \label{eq:GlobalFitEffMaster}
\end{align}
where $I_i$ is the photon flux (photons per time per area) of the illuminations at \SI{1064}{nm} ($i\!=\!0$) and \SI{700}{nm} ($i\!=\!1$), $\Gamma (I_i)$ is the effective decay rate of the charge carriers, \blue{
and $\bar{N}(I_i)$ is the equilibrium number of charge carriers at given intensities $I_i$. 
We then perform a perturbative expansion around the mean intensity $\bar{I_1}$ (see Appendix~\ref{sec:MASTERTheory} for more details), which yields the linearized equation:
\begin{align}
\dot{\delta N}=-\Gamma \delta N +K_1\delta I_1,
\end{align}
where $K_1= K_1(\bar{I_i})$ is the effective cross section for \SI{700}{nm} photons, and $\Gamma= \Gamma(\bar{I_i})$ is the decay rate. Both are a function of the mean intensities $\bar{I_i}$.}
This model predicts a simple pole transfer function 
\begin{align}
\delta \phi_{\mathrm{biref}} &=\frac{2 \eta C^{\rm ext}_{EO} e}{\epsilon_0 A}
    \frac{ g_{\rm DC} }{1+i\omega/\Gamma}
    \delta I_1,\label{eq:TF7}
\end{align}
where
\begin{align}
    g_{\rm DC}
&= \frac{\partial }{\partial I_1} \bar{N}(\bar{I}_i) = \frac{K_1}{\Gamma} \,\,\,\,\,\,\,\,\, {\rm and} \,\,\,\,\,\,\,\,\, \Gamma =\Gamma (\bar{I}_i). \label{eq:gdc_gamma}
\end{align}
Here $C^{\rm ext}_{EO}$ is the measured coating response to external electric fields \cite{Tanioka_2023}, the factor $2$ accounts for both polarizations (fast minus slow), $\eta<1$ is a coupling efficiency describing how much weaker the coating response is to electric fields from the photo-generated charge carrier distribution, and $\epsilon_0$ is the vacuum permittivity.

We can in principle measure the full functions $g_{\rm DC}(\bar{I}_i)$ and $\Gamma (\bar{I}_i)$. For our dataset we were able to get a relatively good empirical fit \blue{across the measured intensity range} for these two functions with just four parameters $\Gamma_0$, $a_0$, $a_1$ and $K_{10}$:
\begin{align}
 \Gamma &=\Gamma_0 + a_0 \bar{I}_0 + a_1 \bar{I}_1, \\ 
    g_{\rm DC}
&= \frac{K_{10}}{\Gamma(\bar{I_0},0)\Gamma(\bar{I_i})} = \frac{K_{10}}{\Gamma_0 \!+\! a_0 \bar{I}_0}  \cdot
\frac{1}{\Gamma_0 \!+\! a_0 \bar{I}_0 \!+\! a_1 \bar{I}_1}, 
\label{eq:gDCMain}
\end{align}

\blue{where the $\Gamma_0$ is the y intercept of the fit, and the two coefficients $a_i$ can be interpreted as cross-sections for photo-induced recombination (or excitation to a non-participating energy level) for the two illumination intensities. $K_{10}$ is an overall gain.} \blue{Similarly, $g_{\rm DC} \Gamma/A = K_{10}/(\Gamma_0+a_0\bar{I_0})/A$ 
is the capture probability for $i\!=\!1$ photons.}

\renewcommand{\arraystretch}{1.4}
\begin{table}[ht]
\centering
{\bf Global Fitting Results}\\
\begin{tabular}{lcr}
\hline
Name & Parameter & Value  \\
\hline

{y intercept}&
$\Gamma_0$ & $-908\pm 39\,\unit{s^{-1}}$ \\
Recomb. cross sections  \\
\;\;\;\;(\SI{1064}{nm}) &
$a_0$& $(1.96\pm 0.06)\times10^{-22}\,\unit{m^2}$\\

\;\;\;\;(\SI{700}{nm}) &
$a_1$& $(4.79\pm 0.15)\times10^{-16}\, \unit{m^2}$\\
Gain &
$\eta K_{10}/A$ & $0.80\pm 0.04\,\unit{s^{-1}}$\\
Capture probability \SI{700}{nm} &
$\eta g_{DC}\Gamma/A$ & \\ \;\;\;
$(\bar{I_0}=2.0\,\unit{MW/m^2})\,$ &&\;\;$6.58_{-0.64}^{+0.71}\times10^{-4}$
 \\\;\;\;
$(\bar{I_0}=2.9\,\unit{MW/m^2})\,$ &&\;\;$3.74_{-0.33}^{+0.36}\times10^{-4}$
\\\;\;\;
$(\bar{I_0}=3.7\,\unit{MW/m^2})\,$ && \;\;$2.71_{-0.23}^{+0.25}\times10^{-4}$
\\
\hline
\end{tabular}
\caption{Result from  global fit to master equation, \blue{including fitting uncertainties}.
$\eta$ is the coupling efficiency of the excited charge carrier, compared to external electric fields, see equation \ref{eq:etaDefExt}. The fit values for $\eta K_{10}/A$ also assume $C^{\rm ext}_{EO}= 1.3\times 10^{-10} \,\, \rm rad/(V/m)$, the measured coating response to external electric fields~\cite{Tanioka_2023}. \blue{Note that the light intensities are converted to photon flux intensities (in units of $\# /\rm{[m^2\cdot s]}$. )}
}
\label{tab:fit_parameters}
\end{table}

Our experiment is only sensitive to the product $C_{EO} K_{10} = C^{\rm ext}_{EO} \eta K_{10}$. Thus we only report values for the combination $\eta K_{10}$.
The results from a global fit for the model parameters are shown in Table~\ref{tab:fit_parameters}. The table also includes values for the capture probabilities $\eta g_{\rm DC} \Gamma/A$ at the \SI{1064}{nm} intensities ($\bar{I}_0$) where we captured our data.

\blue{One additional note on the quality of our empirical linear fit:  The data strongly supports that $\Gamma$ is linear in  $\bar{I}_1$ (\SI{700}{nm}). This can be seen in FIG.~\ref{fig:dcgain_fit}, where the model (dashed lines) is given by equation \ref{eq:gDCMain}. As a result the (positive) quantity $\Gamma_0 \!+\! a_0 \bar{I}_0$ can reasonably be interpreted as decay rate in the absence of \SI{700}{nm} light.
On the other hand we only have data at three different values of $\bar{I}_0$ (\SI{1064}{nm}), and they suggest the true $\bar{I}_0$ dependence might be non-linear outside the small range we observed. As a result, the (negative) y intercept $\Gamma_0$ should not be interpreted as dark decay rate.}

We already mentioned that we can expect the coupling efficiency $\eta$ to be less than one. Our measurement also suggests a lower limit: Since $g_{DC}\Gamma/A$ is the capture probability for photons, it has to be less than one, which in turn means that $\eta$ is bounded below by the largest measured value of $\eta g_{DC}\Gamma/A$ in table \ref{tab:fit_parameters}:
\begin{align}
6.58\times 10^{-4} \times \frac{1.3\times 10^{-10} \,\, \rm rad/(V/m)}{C^{\rm ext}_{EO}}  < \eta < 1.
\label{eq:etarange}
\end{align}
Here we included the explicit dependence of the result in table \ref{tab:fit_parameters} on the previously measured $C^{\rm ext}_{EO}$.

\begin{figure}[ht]
    \centering
    \subfigure[$\;$1064 nm intensity = 2.9 $\rm MW/m^2$.]{%
        \includegraphics[width=0.45\textwidth]
        {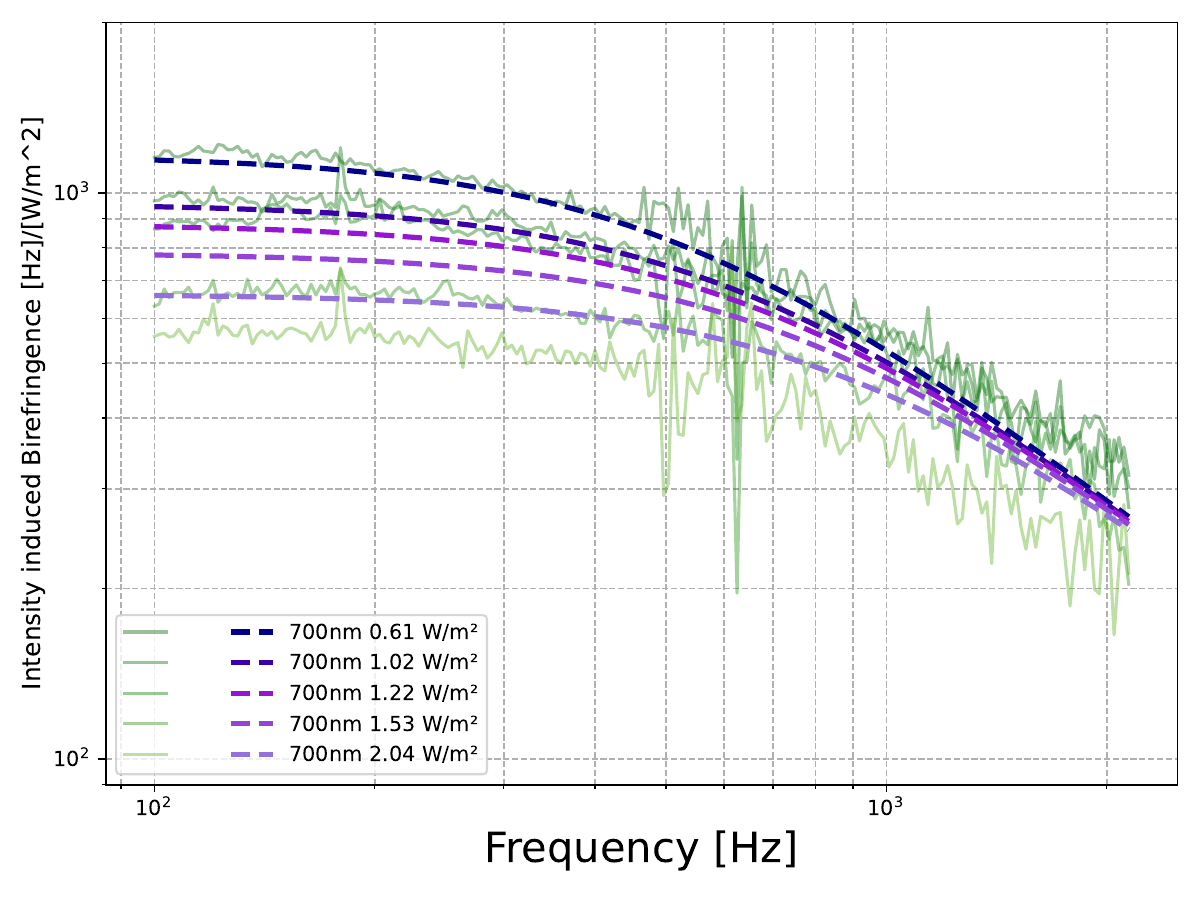}
        \label{fig:0408fit}
    }
    \hfill
    \subfigure[$\;$1064 intensity = 3.7 $\rm MW/m^2$.]{%
        \includegraphics[width=0.45\textwidth]{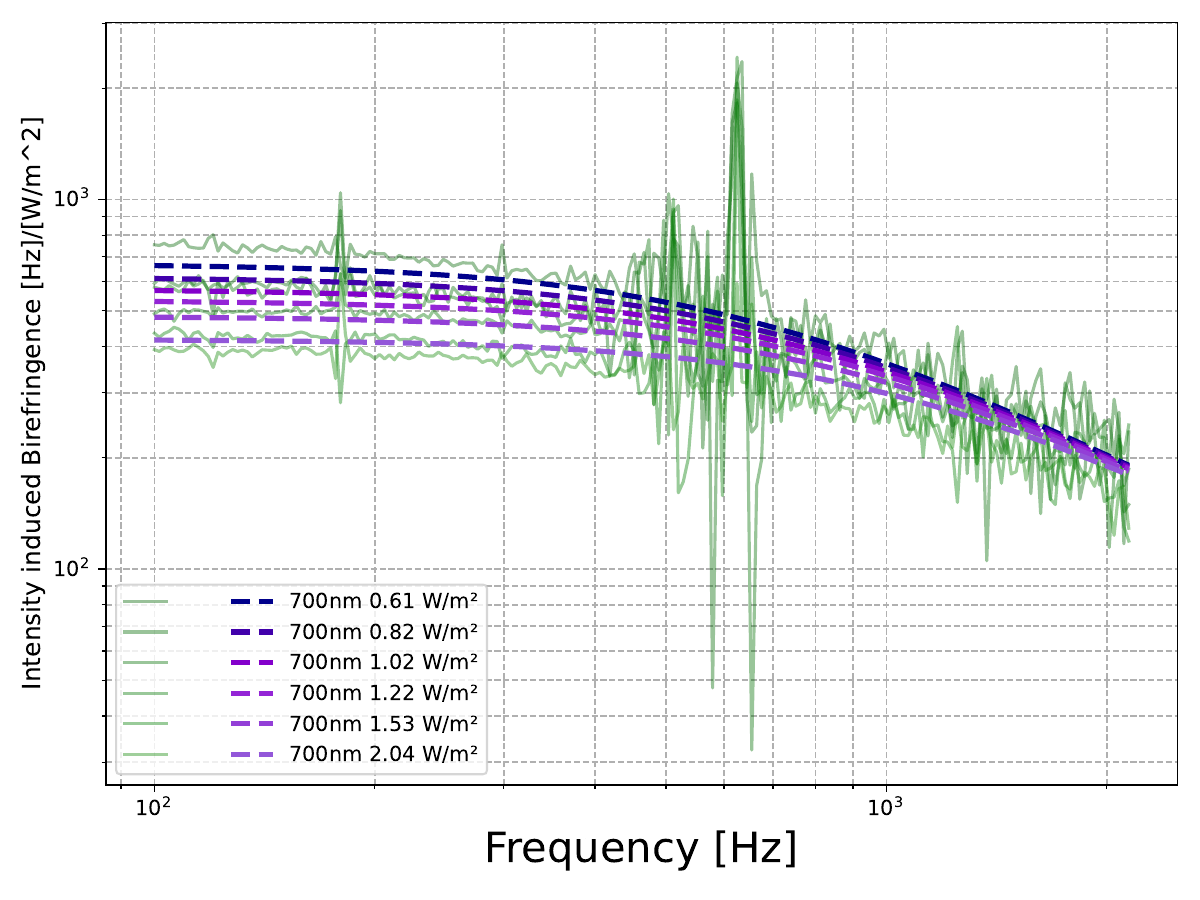}
        \label{fig:0708fit}
    }
    \caption{
    Single global fit for two sets of data with different illumination situations. Solid lines are measured raw data. Dashed lines are calculated from the global model introduced in Appendix ~\ref{sec:MASTERTheory}. This model reproduces the frequency-dependent low-pass shape of the transfer functions and also shows that the DC coupling values and pole frequencies shift with the illumination intensity.}
    \label{fig:globalfit}
\end{figure}

\subsection{Photo-Optic Transfer Function}
While we measured the transfer function from above-band-gap intensity modulation to birefringence observed at the \SI{1064}{nm} carrier wavelength, the master equation model \ref{eq:GlobalFitEffMaster} is symmetric in the indices $i=0,1$. We thus can also predict the transfer function from \SI{1064}{nm} carrier light to birefringence, that is the intensity-to-phase coupling due to the coating. That transfer function
has the same pole frequency $\Gamma/(2 \pi)$, see equation \ref{eq:tf1064}. The strength of that coupling is dependent on an integration constant, which we did not measure in our experiment. However, that strength could be predicted using the pole frequency measured here and tracking the DC behavior of the birefringence as a function of the \SI{1064}{nm} light intensity, as done for instance in \cite{Yu2023}.

\subsection{Generation-Recombination Noise \blue{and Photo-Optic Noise}}

\blue{Two distinct noise sources arise from the light-induced birefringence. First, the master equation~\ref{eq:GlobalFitEffMaster} describes a probabilistic process where charge carriers are generated and recombine individually. This probabilistic process will contribute a fundamental shot noise to the carrier number $N$, which is known in the literature as generation-recombination (GR) noise~\cite{GRN}. Second, intensity fluctuations in the illuminating fields $I_i$ couple to $N$ through the transfer function derived above (Equation~\ref{eq:TF7}), producing an additional noise contribution which we refer to as photo-optic noise. Both noise sources drive fluctuations $\delta N$, which in turn produce fluctuations in the birefringence via the electro-optic effect, and can therefore be expressed as effective displacement noise for the coating. The expected GR noise one-sided displacement power spectral density $S^{\rm 1-s}_{{\delta x}^2,{\rm GR}}$, as well as the photo-optic noise driven by intensity noise $S^{\rm 1-s}_{{\delta x}^2,{\rm PO}}$ can be expressed as following (see appendix~\ref{GRNTheory} for a more detailed derivation):}
\begin{align}
    S^{\rm 1-s}_{{\delta x}^2,{\rm GR}} = 
    &  \left(\frac{\lambda}{4 \pi}
    \frac{C^{\rm ext}_{EO} \, e}{A \, \epsilon_0} \right)^2 \frac{   4 \eta^2 \bar{G} }
    {\Gamma^2 + \omega^2},
    \label{eq:Sdx1sGR1}\\
        S^{\rm 1-s}_{{\delta x}^2,{\rm PO}} = 
    &  \left(\frac{\lambda}{4 \pi}
    \frac{C^{\rm ext}_{EO} \, e}{A \, \epsilon_0} \right)^2 \frac{   
    \left( \eta K_i \bar{I}_i \right)^2 S^{\rm 1-s}_{{\rm RIN_i}^2}   }
    {\Gamma^2 + \omega^2}.
    \label{eq:Sdx1sPO1}
\end{align}
\blue{Here $S^{\rm 1-s}_{{\rm RIN}^2}$ is the relative intensity noise one-sided power spectral density of the illumination $I_i$. $\bar{G}$ is the generation rate at equilibrium, which is equal to the recombination rate $\bar{R}$ at equilibrium. Crucially, they are not necessarily equal to $\Gamma \bar{N}$.}
\blue{Therefore} equation~\ref{eq:Sdx1sGR1} cannot predict the magnitude of the GR noise. \blue{But} it does predict the spectral shape and, assuming $\bar{G}$ is proportional to the incident power, the scaling of GR noise (see appendix \ref{GRNTheory} equation \ref{eq:scaling}):
In words, below the pole frequency $\Gamma/(2\pi)$ the GR noise is {\it white}, scales with power {\it the same way laser shot noise does}, and is {\it independent of the spot size} $A$ for constant power $\bar{P}_0$, as long as the spot radius is larger than the Debye length $\lambda_D$. Remarkably, no other known displacement noise in the observation band of gravitational-wave interferometers is white, making this a signature hallmark of GR noise.


\section{Discussions}\label{sec:discuss}

The observed transfer function can be \blue{described} by modeling the dynamics of \blue{charge carriers trapped somewhere} in the AlGaAs/GaAs coating, \blue{which in turn couple to the coating's birefringence}. Photons absorbed by the coating can excite charge carriers, \blue{and their lifetime} is affected by incident light, with higher intensities shortening the lifetime. The time evolution of the charge carrier density then determines the shape of the transfer function: On time scales shorter than the lifetime, the occupation level is the integral of the illumination intensity, resulting in a $1/f$ transfer function.
On time scales longer than the lifetime, any increase in occupation level due to higher illumination intensities is suppressed by the shortened lifetime of the \blue{carriers}.

The pole frequency is thus determined by the effective carrier lifetime: $f_p=\Gamma/(2\pi)$. Based on our measurements, the effective lifetime of the induced charge carriers is on the order of a few hundred microseconds. This is longer than \blue{the typical carrier lifetime in bulk AlGaAs and GaAs semiconductors~\cite{zarem1989effect,ioffe}, pointing to heterostructure-specific effects. We would expect that the different band gaps of GaAs and AlGaAs would lead to band bending~\cite{doi:10.1021/cr3000626}, forming a 2D potential well with longer lifetimes. Alternatively, DX centers (deep donors) in the AlGaAs barriers could act as metastable traps, affecting the observed lifetimes, similar to what is seen in persistent photoconductivity~\cite{DX,Photoconductivity}.}

Our model further reveals two new noise sources for crystalline coatings (equation~\ref{eq:Sdx1sGR1} and~\ref{eq:Sdx1sPO1} ): the photo-optic noise which is driven by the laser intensity noise, and the GR noise which is an effective shot noise scaling inversely with the laser power below the pole frequency. We are unable to predict the magnitude of those two noise sources because in this experiment we only modulated the LED illumination intensity. Also, the coupling coefficient ($\eta$) remains hard to determine. Finally note that the coupling coefficient of the external electric fields in~\cite{Tanioka_2023} is obtained in the 20-40 kHz frequency band, whereas our range of interest lies below 10 kHz.

Table~\ref{tab:intensities} lists the mean laser power intensity on the test masses of different gravitational-wave detectors, together with the estimated pole frequency of the 
intensity-to-phase transfer function and the GR noise. Note that in all cases the pole frequency is above the observation band, that is, the shortened lifetime due to the high intensity helps to suppress the GR noise and intensity noise coupling. 

\begin{table}[ht]
\centering
\caption{\bf Mean laser intensity and GR pole frequency for  different gravitational-wave detectors.}
\begin{tabular}{lcc}
\hline
Detector & Laser intensity [$\rm kW/cm^2$] & Pole frequency [$\rm kHz$]\\
\hline
aLIGO & 4.5 & 7.4\\
LIGO A+ & 9.1 & 15\\
LIGO A\# & 17 & 28\\
CE & 3.3 & 5.4\\
\hline
\end{tabular}
  \label{tab:intensities}
\end{table}

All the results obtained in this paper are from the same coating sample. We would like to test different samples with identical coatings to make it more convincing that the observed characteristics represent universal properties of AlGaAs/GaAs coatings.

\section{Conclusion}\label{sec:concl}
In this paper, light-induced birefringence is investigated for $\rm Al_{0.92}Ga_{0.08}As/GaAs$ coating under different illumination conditions. By measuring the transfer functions from the illumination to the beating frequency between the s- and p-polarizations,  we observed a frequency dependence of the illumination to birefringence coupling coefficient: flat at lower frequencies and roll-off at higher frequencies. The DC values and pole frequencies are determined by both the outside LED illumination intensity and the carrier intensity. There is a clear trend showing that higher illumination intensities result in lower DC gains and higher pole frequencies for the transfer functions. The measured DC response is consistent with previous research, while the high-frequency roll-off characteristic is observed for the first time.

To describe the measurement results we proposed a master equation model for charge carriers coupling via the electro-optic effect. Overall, our model reproduces the low-pass trend and intensity dependence of the transfer functions. The master equation also predicts GR noise and photo-optic noise. The GR noise is expected to be white below the pole frequency and to scale with power the same way laser shot noise does, independent of the beam spot size.

Further research is required to refine the model. In the current experiment, we were unable to modulate the carrier intensity (1064~nm) and measured only three sets of data with different carrier intensities. As a result we cannot with certainty extrapolate the carrier intensity data much beyond our data range ($I_0$ between $2.0$ and $3.7$~$\rm MW/m^2$).
Additionally, investigating the temperature dependence of induced birefringence would be valuable, as it may provide clues as to whether photon-phonon interactions are involved.

\section{Acknowledgment}\label{sec:ack}
The work in this paper was supported by the National Science Foundation award 
PHY-2513058 and PHY-2309296. We would also like to thank Professors Martin Fejer, Garrett Cole and Steven Penn for many fruitful discussions. This paper was assigned the LIGO DCC number LIGO-P2500676.

\begin{appendices}
\section{A Theoretical Model For Light Induced Birefringence}\label{sec:MASTERTheory}

\subsubsection{Model}

To model the observed birefringence dependence on illumination sources, we assume that the birefringence change $\Delta \phi_{\mathrm{biref}}$ is caused by the local electric field change $\Delta E$ of a charge carrier population trapped somewhere in the coating:
\begin{equation}
\begin{aligned}
    \Delta \phi_{\mathrm{biref}} &=2 C_{EO}  \Delta E =\frac{2 C_{EO}}{\epsilon_0} \Delta \sigma,
    \label{eq:sigmaDef}
\end{aligned}
\end{equation}
with $\epsilon_0$ the vacuum permittivity and $\Delta \sigma$ the change in the carrier area density causing the electric fields across the coating. The electro-optical coupling coefficient $C_{EO}$ encodes the response of the coating to the particular field configuration caused by trapped charge carrier density.
\blue{The factor of $2$ accounts for the fact that the induced birefringence signal receives contributions from both polarizations adding constructively. This anticorrelation between the two polarizations arises from the electro-optical effect in the zincblende crystal structure. The electric field can modify the refractive indices of the fast and slow axes in opposite directions (see Equations (8) and (9) of ref~\cite{Tanioka_2023}). This anticorrelated response was also directly observed experimentally in ref~\cite{Yu2023}, where the two polarizations were measured to have anticorrelated photo-induced frequency fluctuations.}

We are interested in the time evolution and noise of this carrier area density $\sigma(t)$. Thus we express $\sigma(t)$ in terms of the number of participating charge carriers $N(t)$, elementary charge $e$ and an effective area $A$:
\begin{equation}
    \sigma(t) = \frac{e}{A} N(t).
\end{equation}
If the 1064~nm readout radius $w$ is much larger than the Debye length $\lambda_D$ for transverse diffusion, $A$ is simply the 1064~nm readout beam spot area $\pi w^2$. If $w \ll \lambda_D$, $A$ is the carrier transverse diffusion area $\approx \pi \lambda_D^2$. If $w$ and $\lambda_D$ are comparable, $A$ is a diffusion-enlarged spot size.
Since we chose a very large illumination spot, effectively uniform illumination, $A$ does not matter for the transfer function calculation, but it will matter for noise considerations.

The time evolution of the number of charge carriers $N(t)$ is in general governed by a master equation (see for instance ~\cite{GRN}):
\begin{equation}
\begin{aligned}
\dot{N} &=-R(N,I_i)     + G(N,I_i) \equiv F(N,I_i).
\end{aligned}
\label{eq:MainMaster}
\end{equation}
$G(N,I_i)$ is the charge carrier generation rate, and $R(N,I_i)$ is the recombination rate (including excitation to a non-participating state). Both $G$ and $R$ in general depend on the population $N$ itself, as well as the illuminating external optical photon flux $I_i$. In the model we explicitly include the 1064~nm carrier intensity $I_0$, and the 700~nm carrier intensity $I_1$.

To simplify notation, we also introduce the function $F(N,I_i)=G(N,I_i)-R(N,I_i)$ for the right side of the master equation. For constant external fields $I_i$ the system will settle at a mean carrier number $\bar{N}(I_i)$, defined by
\begin{equation}
    F(\bar{N}(I_i),I_i)=0.
    \label{eq:MeanDefinition}
\end{equation}
\blue{
Taking the total derivative $\frac{d}{d I_j}$ and using the chain rule this constraint implies
\begin{equation}
\frac{\partial F}{\partial N} \left(\bar{N}(I_i),I_i\right)  
\frac{\partial\bar{N}}{\partial I_j} \left(I_i\right)  +
\frac{\partial F}{\partial I_j} \left(\bar{N}(I_i),I_i\right) =0
.
\end{equation}
In particular, we are interested in the dependence of $\bar{N}(I_i)$ on $I_1$ (the 700~nm illumination intensity). }We find
\begin{equation}
\frac{\partial\bar{N}}{\partial I_1} \left(I_i\right)  = -
\frac{\frac{\partial F}{\partial I_1} \left(\bar{N}(I_i),I_i\right) }{  \frac{\partial F}{\partial N} \left(\bar{N}(I_i),I_i\right) } .
\end{equation}

In our experiment, the 1064~nm carrier intensity is fixed, i.e., $I_0$ is time independent, while the 700~nm illumination $I_1(t)$ is modulated. We can thus perform a perturbative expansion of the master equation \ref{eq:MainMaster}:
\begin{align}
N(t) &= \bar{N} + \delta N(t), \quad \text{with} \quad |\delta N(t)| \ll \bar{N},\label{perturbation}\\
I_0 &= \bar{I}_0, \\
I_1(t) &= \bar{I}_1 + \delta I_1(t), \quad \text{with} \quad |\delta I_1(t)| \ll \bar{I}_1, \label{perturbationI}
\end{align}
\blue{where $\bar{N}\equiv\bar{N}(\bar{I_i})$. }We thus obtain the linearized equation:
\begin{align}
\dot{\delta N}(t) &= 
\frac{\partial F}{\partial N} (\bar{N},\bar{I}_i) \,\,\, \delta N(t) +\frac{\partial F}{\partial I_1} (\bar{N},\bar{I}_i) \,\,\, \delta I_1(t)
\\&= -\Gamma(\bar{N},\bar{I}_i) \,\,\,\,\, \delta N(t) 
+ K_1(\bar{N},\bar{I}_i) \,\,\,\,\,\,\, \delta I_1(t),
\label{eq:LinearMaster}
\end{align}
with the effective decay rate
\begin{align}
\Gamma(\bar{N},\bar{I}_i) \equiv -
\frac{\partial F}{\partial N} (\bar{N},\bar{I}_i) ,
\label{eq:GammaFirst}
\end{align}
and the effective generation rate
\begin{align}
K_i(\bar{N},\bar{I}_i) \equiv
\frac{\partial F}{\partial I_i} (\bar{N},\bar{I}_i) .
\end{align}
The frequency domain response function is thus given by
\begin{align}
\frac{\delta N}{\delta I_1} &= 
\frac{K_1}{\Gamma +  i \omega} \\ &=  \frac{g_{\rm DC}}{1+i\omega/\Gamma},
\label{eq:tf}
\end{align}
\blue{ where the dependency of $\Gamma$ and $K_i$ on $\bar{N}$ and $\bar{I_i}$ is suppressed.} This corresponds to a simple pole transfer function with pole frequency
\begin{align}
     f_p=\frac{\Gamma}{2\pi},
\end{align}
and DC gain
\begin{align}
    g_{\rm DC}
\equiv \frac{K_1}
     {\Gamma} = \frac{\partial\bar{N}}{\partial I_1},
     \label{eq:gDC}
\end{align}
as we should expect since $\bar{N}$ is the equilibrium carrier number.


\blue{It is useful to Taylor-expand the master equation \ref{eq:MainMaster} up to linear order in $N$ around $\bar{N}(I_i)$, leaving the $I_i$ dependence exact}:
\begin{align}
    \dot{N} = -\Gamma \left(I_i \right) \,\,\left( N - \bar{N}(I_i) \right) + O\left(\left( N - \bar{N}(I_i) \right)^2 \right)
    \label{eq:EffMaster}
\end{align}
\blue{with $\Gamma \left(I_i \right)\equiv\Gamma \left(\bar{N}(I_i),I_i \right)$.} 
Our experiment directly measures \blue{$\Gamma(\bar I_i)$} and the derivative \blue{$\frac{\partial}{\partial I_1}\bar{N}(\bar I_i) = g_{\rm DC}(\bar I_i) $}, up to an uncertain coupling coefficient $C_{EO}$. 
\blue{Note that the effective decay rate $\Gamma \left(N,I_i \right)$, defined via
\begin{align}
\Gamma(N,I_i) \equiv  \frac{\partial R}{\partial N} (N,I_i) - \frac{\partial G}{\partial N} (N,I_i),
\label{eq:GammaFirstb}
\end{align}
is different from the individual charge carrier decay rate $\gamma(N,I_i)$, which can be defined via 
\begin{align}
R(N,I_i)=\gamma({N},I_i) N.
\label{eq:Def_gamma}
\end{align}
}



Finally, we note that the master equation~\ref{eq:MainMaster} is symmetric in the fields $I_0$ and $I_1$. Thus, by swapping the indices 0 and 1, we find the shape of the intensity-to-phase coupling of the coating for the \SI{1064}{nm} light:
\begin{align}
\frac{\delta N}{\delta I_0} &= 
\frac{K_0}{\Gamma +  i \omega}. 
\label{eq:tf1064}
\end{align}
This \SI{1064}{nm} transfer function will have the same pole frequency as the measured \SI{700}{nm} light, see eq. \ref{eq:GammaFirst}. While we did not measure the magnitude of this coupling $K_0$, it could be predicted by just tracking the DC behavior of the birefringence.

\subsubsection{Fitting}

To fit our dataset we fit the two measurable functions $\Gamma(\bar{I}_i)$ and $g_{\rm DC}(\bar{I}_i)$ with a total of four parameters. First, we fit
$\Gamma(\bar{I}_i)$ as a linear function in both $\bar{I}_0$ (the carrier photon flux) and $\bar{I}_1$ (the illumination photon flux), requiring a y-intercept and two slopes: $\Gamma_0$, $a_0$ and $a_1$.
Finally, we need one more overall gain $K_{10}$ to fit $g_{\rm DC}(I_i)$ entirely empirically as:
\begin{align}
 \Gamma &=\Gamma_0 + a_0 \bar{I}_0 + a_1 \bar{I}_1 ,\\ 
    g_{\rm DC}
&= \frac{K_{10}}{\Gamma(\bar{I_0},0)\Gamma(\bar{I_i})} = \frac{K_{10}}{\Gamma_0 \!+\! a_0 \bar{I}_0}  \cdot
\frac{1}{\Gamma_0 \!+\! a_0 \bar{I}_0 \!+\! a_1 \bar{I}_1}. \label{eq:Ansatz}    
\end{align}

Since $\Gamma(\bar{I}_i)$ is an effective decay rate, the 3 parameters $\Gamma_0$, $a_0$ and $a_1$ look like effective spontaneous decay rate and two effective cross sections for photo-induced recombination.
These effective terms are the difference between recombination and generation. Furthermore there is no expectation that this fit is accurate down to zero photon flux, and a negative $\Gamma_0$ is not a contradiction with the model.
Following the same analogy, $K_1 = g_{\rm DC} \Gamma = K_{10}/\Gamma(\bar{I}_0,0)$ is the effective capture cross section for $i\!=\!1$ photons, see equation \ref{eq:LinearMaster}. 

In equation \ref{eq:sigmaDef}, we do not know the true electro-optical coupling coefficient $C_{EO}$ for the particular field configuration caused by trapped charge carrier density we are dealing with. We only know the coefficient for external fields, $C^{\rm ext}_{EO}= \SI{1.3e-10}{rad/(V/m)}$, measured in  \cite{Tanioka_2023}. We expect the relevant coupling $C_{EO}$ to be somewhat smaller than $C^{\rm ext}_{EO}$, so we can write
\begin{equation}
    C_{EO}=\eta C^{\rm ext}_{EO},
\label{eq:etaDefExt}
\end{equation}
with $\eta<1$, a coupling efficiency due to the geometry of the electric fields from the charge carrier distribution. Our transfer function experiment cannot distinguish changes in $\eta$ from changes in $K_{10}$, so we only report values for $\eta K_{10}$ for the fitting results presented in Table~\ref{tab:fit_parameters}. 
However, knowledge of $\eta$ will become important for GR noise estimation presented below.

We can also integrate \blue{the functional dependence of} our fit (equation \ref{eq:Ansatz}) and get an explicit form for the $\bar{N}(I_i)$ and therefore for the master equation in terms of our fitted parameters
\begin{align}
    \dot{N} = -\Gamma \left( N 
    - \frac{K_{10}}{a_1 \Gamma_1}  \left( \ln{\Gamma} -\ln{\Gamma_1} \right)  - \bar{N_0}  \right),
\end{align}
where we introduced the abbreviation $\Gamma_1 = \Gamma(\bar{I_0},0)$, and as expected we cannot determine the integration constant $\bar{N}_0(I_0)$.

\section{GR Noise and Photo-Optic Noise}
\label{GRNTheory}

To understand the noise implications of our model we return to the most general form of the master equation \ref{eq:MainMaster}, written in terms of generation terms $G$ and recombination terms $R$:
\begin{equation}
\begin{aligned}
\dot{N} &=-R(N,I_i) \,
    + G(N,I_i) .
\end{aligned}
\end{equation}
The fluctuations in $N$ are fundamentally due to two processes:
\begin{enumerate}
    \item 
The generation $G$ and recombination $R$ each are probabilistic processes that can only generate or destroy a whole charge carrier. Thus, each contributes a shot noise ($\delta R$ and $ \delta G$ respectively) with (1-sided) amplitude spectral density equal to $\sqrt{2 G(\bar{N},I_i)} \equiv \sqrt{2 \bar{G}}$. The literature refers to this noise as Generation-Recombination noise, abbreviated as GR noise. To calculate the GR noise for our situation, we can follow the description of GR noise in semiconductors outlined in~\cite{GRN}.
\item Intensity fluctuations in the drive fields $I_i$ will couple to the birefringence with the transfer function derived in appendix~\ref{sec:MASTERTheory} (equation \ref{eq:tf} and \ref{eq:tf1064}). We will refer to this coupling as photo-optic noise.
\end{enumerate}

First, we note that we can write the (1-sided) amplitude spectral density for the source terms $\delta R$ and $\delta G$ as
\begin{align}
    \sqrt{2 G(\bar{N},I_i)} =    \sqrt{2 R(\bar{N},I_i)} \equiv
    \sqrt{2 \bar{G}},
    \label{eq:ASD}
\end{align}
\blue{where we assumed $\bar{G}=\bar{R}$ (equilibrium).}
Next, to calculate GR noise, we can again perform a perturbative expansion
\begin{align}
N&=\bar{N}+\delta N, \\
I_0 &= \bar{I}_0 + \delta I_0 , \\
I_1 &= \bar{I}_1 + \delta I_1,
\end{align}
and explicitly add the driving noise terms $\delta R$ and $\delta G$:
\begin{align}
    \dot{\delta N} = 
&-R    \,\,\,\,\,\,\,\, + G     \,\,\,\,\,\,\,\,\, - \delta R + \delta G \\    
    =&- \Gamma  \delta N +
    K_i \delta I_i - \delta R + \delta G,\label{dsigmadt6}
\end{align}
We thus find
\begin{align}
    {\delta N} = 
    &  \frac{   
    K_i \delta I_i - \delta R + \delta G}
    {\Gamma +i \omega}.
    \label{dsigmadt7}
\end{align}
We can now write the 1-sided power spectral density for fluctuations $\delta N$ as
\begin{align}
    S^{\rm 1-s}_{{\delta N}^2} = 
    &  \frac{   
    \left( K_i \bar{I}_i \right)^2 S^{\rm 1-s}_{{\rm RIN_i}^2}  + 4 \bar{G} }
    {\Gamma^2 + \omega^2},
    \label{dsigmadt9}
\end{align}
where $S^{\rm 1-s}_{{\rm RIN_i}^2}$ is the 1-sided power spectral density of the relative intensity noise for the $I_i$ field, and the rate power spectral density of $\delta R$ and $\delta G$ are given by the square of equation \ref{eq:ASD}.

We can now cast this into an effective 1-sided displacement noise power spectral density for the coating
\begin{align}
    S^{\rm 1-s}_{{\delta x}^2} = 
    &  \left(\frac{\lambda}{4 \pi}
    \frac{C^{\rm ext}_{EO} \, e}{A \, \epsilon_0} \right)^2 \frac{   
    \left( \eta K_i \bar{I}_i \right)^2 S^{\rm 1-s}_{{\rm RIN_i}^2}  + 4 \eta^2 \bar{G} }
    {\Gamma^2 + \omega^2},
    \label{eq:Sdx1s}
\end{align}
where we also reintroduced $\eta$ and the measured coupling for external electrical fields $C^{\rm ext}_{EO}$ via $C_{EO}=\eta C^{\rm ext}_{EO}$. 
\blue{Equation~\ref{eq:Sdx1s} can be split into GR noise and photo-optic noise: }
\begin{align}
        S^{\rm 1-s}_{{\delta x}^2,{\rm GR}} = 
    &  \left(\frac{\lambda}{4 \pi}
    \frac{C^{\rm ext}_{EO} \, e}{A \, \epsilon_0} \right)^2 \frac{   4 \eta^2 \bar{G} }
    {\Gamma^2 + \omega^2},
    \label{eq:Sdx1sGR}\\
        S^{\rm 1-s}_{{\delta x}^2,{\rm PO}} = 
    &  \left(\frac{\lambda}{4 \pi}
    \frac{C^{\rm ext}_{EO} \, e}{A \, \epsilon_0} \right)^2 \frac{   
    \left( \eta K_i \bar{I}_i \right)^2 S^{\rm 1-s}_{{\rm RIN_i}^2}   }
    {\Gamma^2 + \omega^2}.
    \label{eq:Sdx1sPO}
\end{align}

{\bf Noise Scaling:} We expect the GR noise to be spatially correlated up to the Debye length $\lambda_D$ due to transverse diffusion, but uncorrelated across the coating for larger separations.
Note that for fixed intensities $I_i$ both $K_i$ and $\bar{N}$ are proportional to the effective area A.
This guarantees that for fixed intensities $I_i$ 
photo-optic noise is independent of $A$, while for the GR noise we have
\begin{equation}
    S^{\rm 1-s}_{{\delta x}^2,{\rm GR}} \propto \frac{1}{A} \,\,\,\, {\rm for~fixed~intensities}~I_i.
\end{equation}


The frequency band of interest for gravitational-wave detectors is roughly \SI{10}{Hz} to \SI{1}{kHz}. For typical operating powers of gravitational-wave detectors this frequency band lies below the pole frequency and we have $\Gamma \approx \Gamma_1 \approx a_0 \bar{I}_0$.
To predict a power scaling, we have to {\it assume } that the generation rate $\bar{G}$ is proportional to the total power $\bar{P}_0\!=\!A \bar{I}_0$ on the optic. Here $A$ is the beam spot area, which for gravitational-wave detectors is much bigger than the diffusion area. 
Thus we find
\begin{equation}
    S^{\rm 1-s}_{{\delta x}^2,{\rm GR}} \propto \frac{1}{A \bar{I}_0} = \frac{1}{\bar{P}_0},
    \label{eq:scaling}
\end{equation}
where $\bar{P}_0$ is the total power on the coating (the interferometer arm power). In words, below the pole frequency the GR noise is {\it white}, scales with power {\it the same way laser shot noise does}, and is {\it independent of the beam spot area} $A$ for constant power $\bar{P}_0$.

\section{Calibration}\label{sec:scheme}

\begin{figure}
\begin{center}
\includegraphics[height=0.21\textheight]
{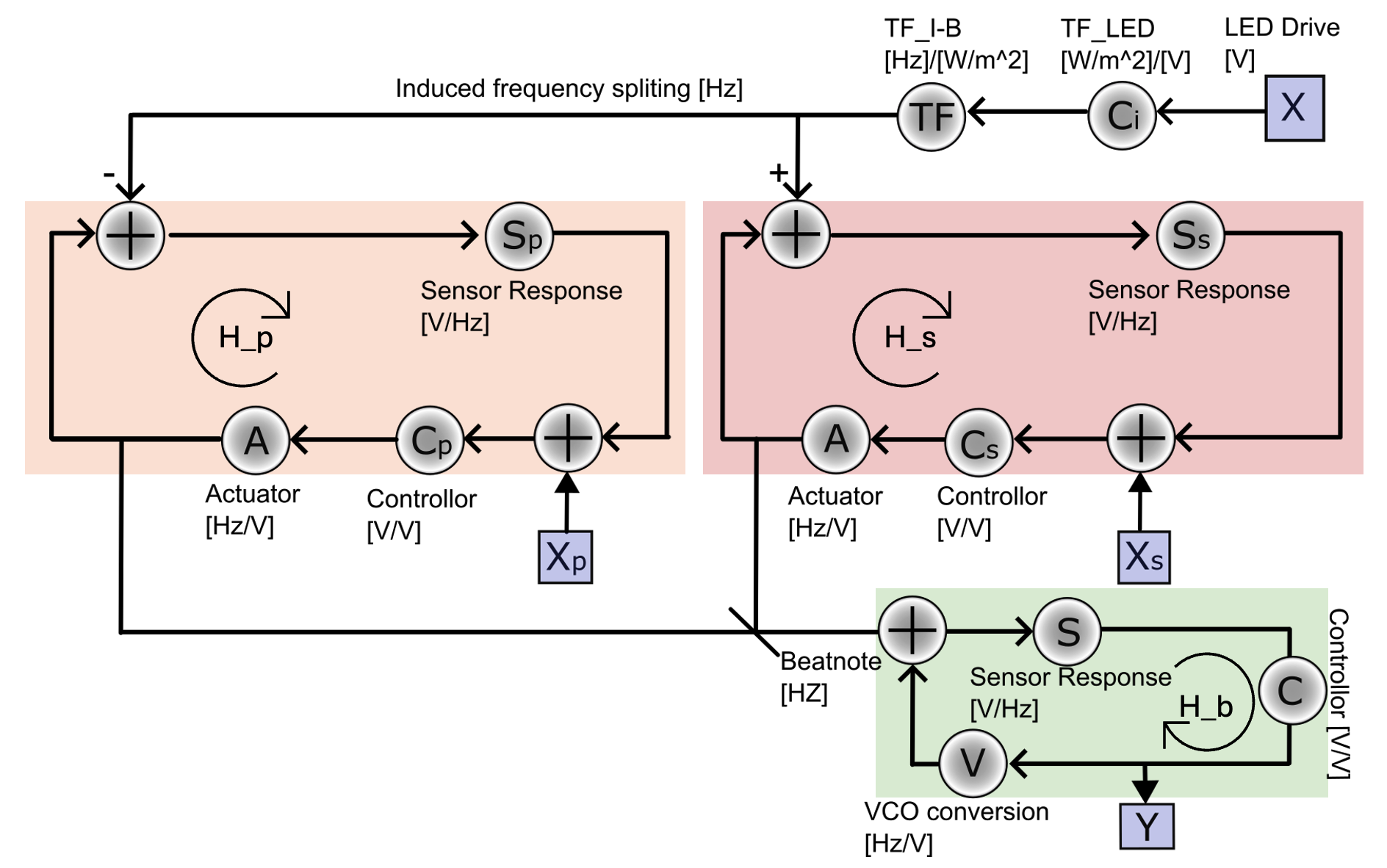}
\caption{Block diagram of the experimental setup. Color-coded are the s- and p- polarization control loops (orange and red), as well as the PLL loop (green) tracking the beat note. \blue{The + and - signs on the induced frequency splitting in the two polarization locking loops indicate that the illumination affects the resonant frequencies of the two polarizations in opposite directions.}}
\label{fig:TF_measurement}
\end{center}
\end{figure}

In order to measure the LED illumination to  birefringence transfer functions we locked both the s- and p- modes of the cavity using Pound-Drever-Hall (PDH) control loops. The beat signal of the two laser polarizations was fed into a phase-locked loop (PLL) to lock the output of a voltage-controlled oscillator (VCO) to the beat note signal, as shown in FIG.~\ref{fig:TF_measurement}. The control signal sent to the VCO is  a direct measure of the beat note frequency.

\subsubsection{PDH Locking Loops for the Two Polarizations}
The laser locking sensing responses $S_s$ and $S_p$ can be obtained from the error signal in the PDH locking process. The slope of the error signal linear range corresponds to the sensing gain. It varied between $\SI{6}{V/GHz}$ and $\SI{18}{V/GHz}$, depending on the laser input power. The measurement was repeated for every data run.

\subsubsection{Transfer Functions}

The LED drive ($ X$) is first converted to the LED light intensity and then induces a birefringent frequency shift for the two orthogonal modes, in a different direction. $ H_p$ and $ H_s$ in the diagram denote the locking loop gains of two orthogonal modes. $ H_b$ denotes the gain of the PLL. The transfer function from illumination to birefringence, which is written as \blue{$\rm TF$} in the block diagram, can then be calculated from the following measurement results.

We first inject an excitation signal from the laser locking error signal node ($X_p$ or $X_s$), and then measure the transfer function from the error signal node to the VCO control signal $Y$. Let us take the p-mode locking loop as an example: the transfer function $\frac{Y}{X_p}$ can be calculated by the following when the cavity noise is relatively small:

\begin{align}
    \frac{Y}{X_p}=&\frac{1}{1-H_p}\times C_p \times A \times \frac{1}{1-H_b}\times S\times C,\label{eq:1Xp_Y}\\
    =&\frac{H_p}{1-H_p}\times \frac{1}{S_p}\times \frac{1}{1-H_b}\times S\times C,\label{eq:2Xp_Y}
\end{align}  
in which $C_p$ and $S_p$ are the controller and sensor response of the p-mode locking loop; $A$ is the laser actuator response and $C$ is the PLL controller.

The second step is to measure the transfer function from the LED drive $X$ to VCO control signal $Y$:
\begin{align}
 \frac{Y}{X}= &C_i\times{\rm TF}\times [\frac{H_s}{1-H_s}-(-\frac{H_p}{1-H_p})]\times \frac{1}{1-H_b}\times S\times C,
\\
      =&C_i\times {\rm TF}\times (\frac{H_s}{1-H_s}+\frac{H_p}{1-H_p})\times \frac{1}{1-H_b}\times S\times C\label{eq:1X_Y},
\end{align}      
where $C_i$ indicates the LED conversion from drive voltage to the illumination intensity and $S$ is the sensor response of PLL.
Now with equation~\ref{eq:2Xp_Y} and equation~\ref{eq:1X_Y}, we find that:
\begin{equation}
    {\rm TF}  =\frac{1}{C_i}\times\frac{Y}{X} \times \frac{1}{({S_s}\frac{Y}{X_s}+ {S_p}\frac{Y}{X_p})   }\label{eq:I3}.
\end{equation} 
In the case where $H_p\gg1$, $H_s\gg1$, $S_p \approx S_s$ we then find 
\begin{equation}
     {\rm TF}  =\frac{1}{S_p}\times \frac{X_p}{Y}\times\frac{1}{2C_i}\times \frac{Y}{X} \label{eq:I2}.
\end{equation} \label{eq:I}

\begin{figure}[ht]
    \centering
    \begin{minipage}[t]{0.2\textwidth}
        \centering
        \subfigure[$\;$700 nm LED calibration]{%
            \includegraphics[width=\textwidth]{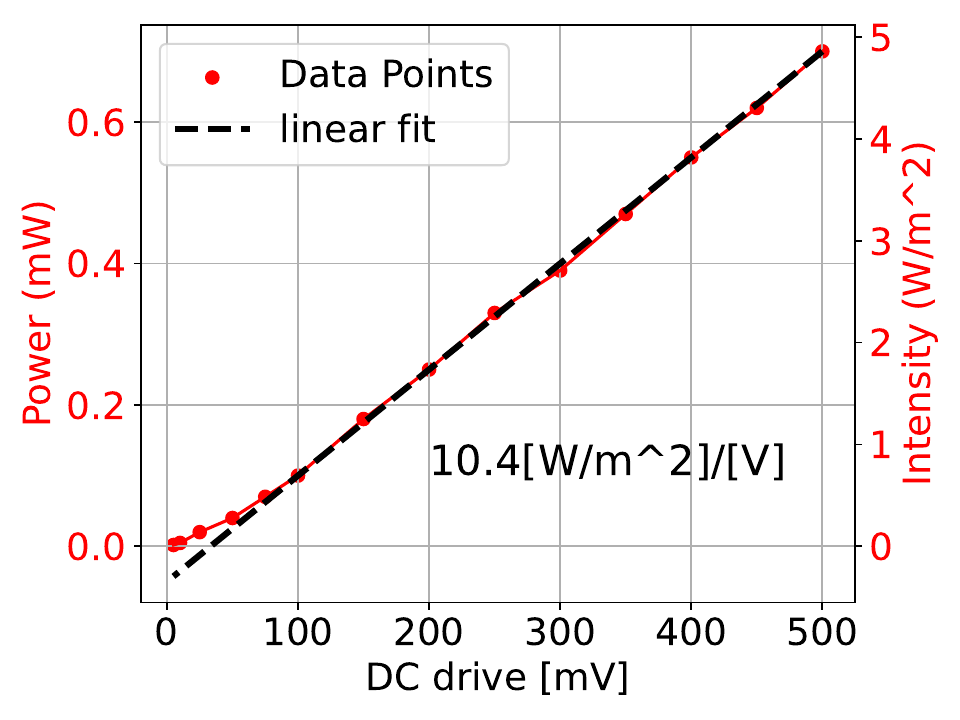}
            \label{fig:700nmLED}
        }
    \end{minipage}%
    \hfill
    \begin{minipage}[t]{0.22\textwidth}
        \centering
        \subfigure[$\;$430 nm LED calibration]{%
            \includegraphics[width=\textwidth]
            {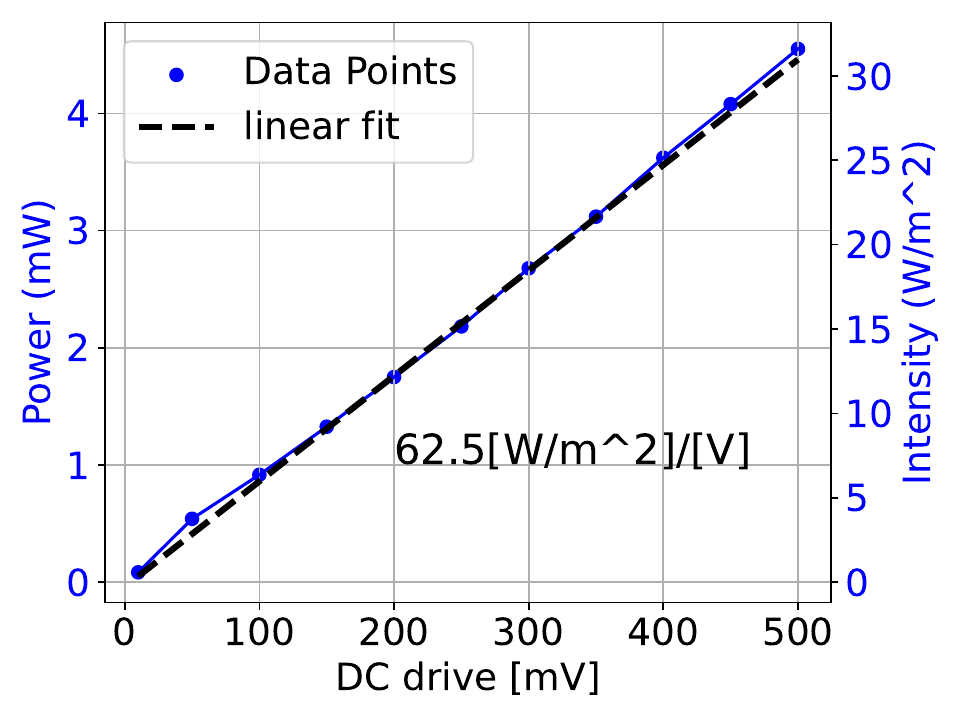}
            \label{fig:400nmLED}
        }
    \end{minipage}
    \caption{LED calibration. For the 700 nm LED, we have: $\rm 10.4 \,(W/m^2)/V$ and for the 430 nm LED, we have: $\rm 62.5 \,(W/m^2)/V$.}
    \label{fig:TF_LED}
\end{figure}
\subsubsection{LED Illumination Calibration}

\begin{figure}
\begin{center}
\includegraphics[height=0.2\textheight]{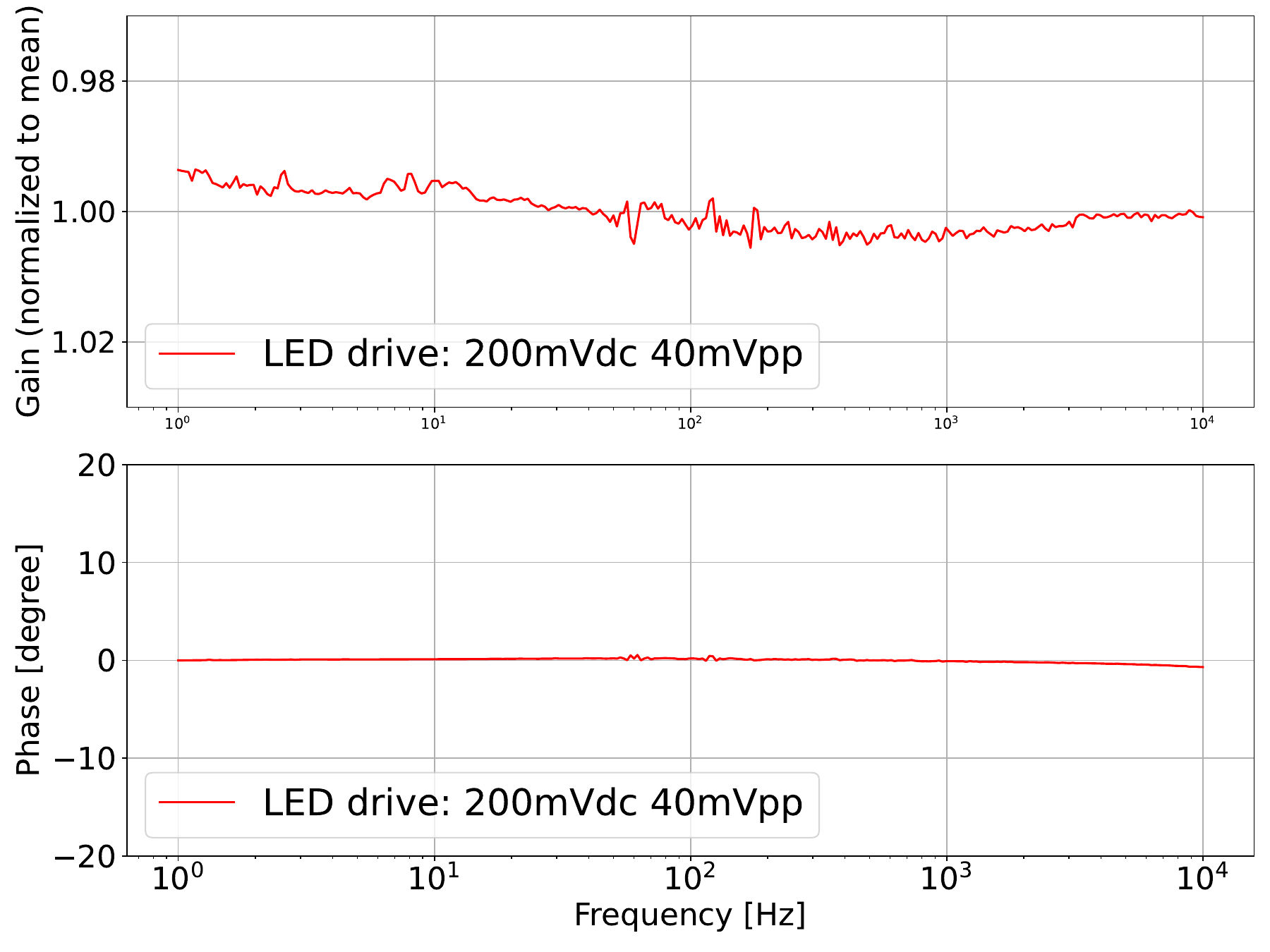}
\caption{Frequency response of the 700 nm LED measured with a drive source of 200 mV DC voltage and 40 mV peak-to-peak modulation from 1 to 10 kHz. The gain value is  normalized to mean and the fluctuations are less than $\pm1\%$. Crucially, the transfer function is flat up to \SI{10}{kHz} without any roll-off pole.}
\label{fig:LED_TF}
\end{center}
\end{figure}

The LED conversion from drive voltage to illumination intensity is shown in Fig.~\ref{fig:TF_LED}.
The external illumination sources at 700~nm and 430~nm were incident under approximately 45 degrees to the mirror surface. Throughout this paper we report the illumination intensity without correcting for the angle cosine or for the effective coating reflectivity at that angle and wavelength. We also measured the LED's frequency response using a photodetector that collects scattered light, see Fig.~\ref{fig:LED_TF} for an example. The gain fluctuations were within $\pm1\%$.

\end{appendices}

\bibliographystyle{unsrt}
\bibliography{biref_paper}

\end{document}